\documentclass{llncs}

\usepackage{amsmath} 

\usepackage{units}
\usepackage{graphicx}

\usepackage[usenames]{color} 
\usepackage[ruled,vlined,linesnumbered]{algorithm2e} 

\newcommand{\ie}{i.e.}

\definecolor{Blue}{rgb}{0,0.16,0.90}
\definecolor{Red}{rgb}{0.90,0.16,0}
\definecolor{DarkBlue}{rgb}{0,0.08,0.45}
\definecolor{ChangedColor}{rgb}{0.9,0.08,0}
\definecolor{CommentColor}{rgb}{0.2,0.8,0.2}
\definecolor{ToDoColor}{rgb}{0.1,0.2,1}

\def \CalA {\mathcal{A}}
\def \CalD {\mathcal{D}}
\def \CalH {\mathcal{H}}
\def \CalP {\mathcal{P}}

\title{An Oblivious Password Cracking Server}

\author{Aureliano Calvo\inst{1} 
  \and Ariel Futoransky\inst{1}
  \and Carlos Sarraute\inst{1,2}}

\institute{CoreLabs Research Center, Buenos Aires, Argentina
\and 
ITBA (Instituto Tecnol\'ogico de Buenos Aires)
}

\begin{document}

\maketitle
\begin{abstract}
Building a password cracking server that preserves the privacy of the queries 
made to the server is a problem that has not yet been solved.
Such a server could acquire practical relevance in the future:
for instance, the tables used to crack the passwords could be calculated, stored and
hosted in cloud-computing services, and could be queried from 
devices with limited computing power.

In this paper we present a method to preserve the confidentiality of
a password cracker---wherein the tables used to crack the passwords
are stored by a third party---by combining Hellman tables and Private
Information Retrieval (PIR) protocols. 
We provide the technical details of this method,
analyze its complexity,
and show the experimental results obtained with our implementation.
\end{abstract}

\section{Introduction}

Suppose that you're a hacker (or pentester) attacking a sensitive computer network,
and that you've gained access to a list of password hashes.
You need to retrieve the corresponding passwords to carry on with the attack...
unfortunately, you don't have access to your Rainbow tables 
(because the computing device used to carry out the attack has limited computing power
and/or memory, for example because you're using a smartphone).

This is when you need a password cracking server, that will
provide access to the relevant parts of the Rainbow tables.
But of course you don't want to reveal to the server 
(that we suppose is managed by a third party) which passwords you are trying to crack.
This is the problem that we tackle in this paper: 
to build an ``oblivious password cracking server'' that contains
tables of passwords and hashes, and that the users can query 
without revealing which pairs of passwords and hashes they are interested in.

The paper is structured as follows.
In Section~\ref{sec:preliminaries} we introduce some background
on the ideas that we used.
Section~\ref{sec:solution} provides an overview of the solution
that we propose for this problem.
In Section~\ref{sec:algorithms-detail} we get into the technical details
of the algorithms, in particular about the construction of the tables.
Section~\ref{sec:complexity} deals with the PIR protocols
and provides computations of their complexity.
In Section~\ref{sec:experimental-results} we show experimental results
obtained with our prototype in Python.
We conclude the paper with ideas for future work.

\section{Preliminaries} \label{sec:preliminaries}

Before describing the proposed solution, we give
a brief background on the ideas that we use: 
hash reversing tables based on time-memory trade-offs,
and Private Information Retrieval (PIR) schemes
to query the database server.

\subsection{Hash Reversing Tables}\label{sec:intro-hrt}

A common approach in computer systems that rely on passwords for authentication
is to store a cryptographic hash of the password. 
This approach is vulnerable to attacks based on precomputed tables for reversing 
the cryptographic hash function.

Martin Hellman proposed in 1980 a time-space
trade-off to reverse one-way functions \cite{paper:h-80}, 
and thus make such precomputed tables more practical.
The insight of Hellman was to compute chains
of hashes and passwords, and to store only the beginning
and end of each chain.

Ron Rivest then proposed (in 1982) an improvement over
the Hellman tables \cite{inbook:r-82}. The idea was
to use some (distinguished) images as chain ends in order to reduce the
number of table lookups.
As we will discuss in Sections \ref{sec:hash-reversing} and \ref{sec:comparison},
these tables are particularly suited for the purpose sought in this work.

In 2003, Philippe Oeschlin proposed a new improvement over the Hellman
tables \cite{paper:o-03}. Instead of using a single reduction function
for each chain, it would use a different one for each step in the
chain. Although the Rainbow tables are faster than the Hellman tables
in the general case, this is not true in the specific case of querying the database
using a PIR protocol.
We will discuss the details in Section \ref{sec:hash-reversing}.

\subsection{Private Information Retrieval} \label{sec:PIR}

In this work, we are interested in the case of a single database---which
stores the hash reversing tables.
A single-database Private Information Retrieval (PIR) scheme is a
game between two players: a user and a database. The database holds
some public data (for concreteness, an $n$-bit string). The user
wishes to retrieve some item from the database (such as the $i$-th
bit) without revealing to the database which item was queried (\ie,
$i$ remains hidden) \cite{paper:pir_survey}.

A PIR scheme usually consists of 5 steps:
\begin{enumerate}
\item Query generation (happens in the client)
\item Query transmission
\item Query processing (happens in the server)
\item Response transmission
\item Response decoding (happens in the client)
\end{enumerate}

In the rest of the paper, we will note 
$O_{C}$ the client processing complexity (steps 1 and 5), 
$O_{S}$ the server processing complexity (step 3) and 
$O_{T}$ the transfer complexity (steps 2 and 4). 

It is important to note that $O_{S}$ has to be at least $O(n)$ 
to assure that no information is leaked to the server \cite{paper:pir_survey}. 
More details about current PIR schemes are given in Section~\ref{sec:PIR-details}.

\section{Our Proposed Solution} \label{sec:solution}

\begin{figure}[ht]
\centering
\includegraphics[width=0.9 \linewidth]{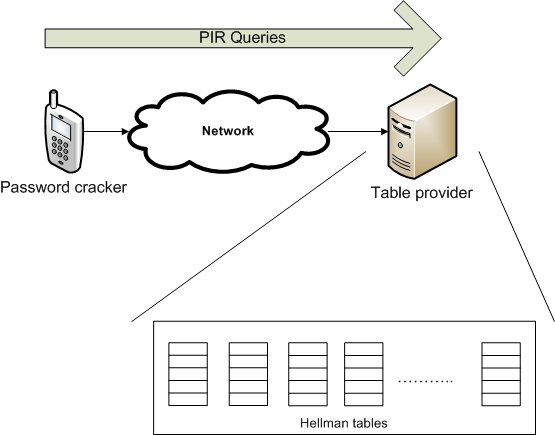}
\caption{Scheme of the proposed solution.} \label{fig:scheme}
\end{figure}

We give in this section an overview of the solution.
The fundamental idea is to store Hellman tables with
distinguished end-points in a series of databases accessible using
a PIR protocol (see Figure~\ref{fig:scheme}). 
When a user of this system
attempts to find the password corresponding to a given hash, she makes a series of
PIR queries to retrieve the beginning and end of
the chains corresponding to the hash being reversed.

Let $\CalP$ be the space of passwords, $\CalH$ the space of hashes,
and $H : \CalP \rightarrow \CalH$ a one-way function used to transform 
a plain-text $p \in \CalP$ into a hash $ h \in \CalH$.
We use $\CalD \subseteq \CalP$ to denote the set of passwords 
we are interested in,
and $N$ for the number of passwords in $\CalD$. 
In order to quickly find in $\CalD$ the password corresponding to a given hash 
with probability $\alpha$, 
we compute and store in the server a set of $M$ Hellman
tables with distinguished end-points, using $M$ different reduction
functions, and whose chain length is on average $M$.
The parameter $M$ depends on $\alpha$ and on the size of $\CalD$,
and is in the order of $N^{1/3}$
(see Section~\ref{sub:distinguished end-points} for details). 

Once the tables are generated, in order to calculate the inverse of
the hash $h \in \CalH$, the corresponding end-point for each reduction function
$r_{i}$ is calculated and looked up in the tables. For each end-point
found in a table, the chain is searched through looking for the preimage of
$h$. If the search is successful, the corresponding password 
$p$ such that $h = H(p)$ is found.

Since PIR protocols are computationally expensive,
our solution is to make a single query to the database for each table. 
A query in a PIR protocol retrieves a set of consecutive bits
from the database. In order to make a single PIR query per table,
the chains are stored in a closed hash table\footnote{In a closed hash table,
each bucket holds at most one entry.}
sorted by the chain end-point.
If a collision between the ends of two chains in the same table
is found, one of the chains is discarded and
another chain is calculated.
We chose to discard colliding chains because
we have to do exactly one PIR query per table to assure
that the server does not gain information regarding the password.
This was an important design decision, and is one of the contributions
of this work.

To ensure that the table can be calculated
in the same running order (as the regular Hellman tables with distinguished
end-points), each hash table is given $\beta M$ slots.
The parameter $\beta > 1$ is used to expand the hash table,
in order to have enough spare buckets and avoid collisions while generating the table.

\section{The Algorithms in Detail} \label{sec:algorithms-detail}

\subsection{Generating the Tables} \label{sec:hash-reversing}

Algorithm \ref{algo:tables_calculation} shows
the procedure used by the table provider to calculate the Hellman
tables with distinguished end-points.

\begin{figure} [ht]
\vspace{-0.5 cm}
\begin{algorithm}[H]
  \caption{Calculate tables}
\label{algo:tables_calculation}%
  \DontPrintSemicolon

  \KwIn{$\alpha, M$}

  \KwOut{tables}

  tables $\gets$ empty\_list() \;
  \For{$i \in [0,M)$}{
    table $\gets$ new array(size = $\alpha \cdot M$, default = EMPTY\_ENTRY) \;
    chain\_count $\gets 0$ \;
    chain\_index $\gets 0$ \;
    \While{\emph{chain\_count} $ < M $}{
      end $\gets$ calculate\_end\_point(redfun(index), password\_for(chain\_index)) \;
      bucket\_idx $\gets$ bucket\_for(end) \;
      \If{\emph{table[bucket\_idx] == EMPTY\_ENTRY}}{
        bucket\_idx = (chain\_index, end) \;
        chain\_count += $1$ \;
      }
      chain\_index += $1$ \;
    }
    tables.append(table) \;
	}
  return tables \;
\end{algorithm}
\vspace{-0.5 cm}
\end{figure}

In our implementation, the generation of the tables depends on:
(i) the alphabet used by the passwords, 
(ii) the length of the passwords,
and (iii) the desired probability $\alpha$ of cracking the passwords.
Of course, the tables also depend on the hash function $H$ to be reversed,
on the reduction functions used to create the chains,
and on the parameter $\beta$.

In the following subsections, we give the relevant details 
of the different versions of time-space trade-offs used for hash reversing.
We also discuss why the Hellman tables with distinguished end-points
were most suited for this application.

\subsubsection{Hellman Tables.}

In \cite{paper:h-80} Hellman proved that by making a table
of size $N^{2/3}$ a one-way function $H : \CalP \rightarrow \CalH$ 
can be reversed in $O(N^{2/3})$
operations where $N$ is the size of the function domain.

Let $M$ be both the number of reduction functions and the
number of steps of a chain. The probability of success of finding the preimage
of a hash obtained from the table domain can be estimated as 
$\alpha = 1-e^{-\frac{M^{3}}{N}}$ \cite{paper:o-03}. 
This means that in order to have a success probability
$\alpha$, $M$ should be computed as: 
\begin{align} \label{eq:m}
M = -\sqrt[3]{ \ln(1-\alpha) \cdot N}
\end{align}

The scheme consists in chaining $M$ results by defining $M$
reduction functions $r_{i} : \CalH \rightarrow \CalP$.
Each function $r_{i}$ is used to generate
$M$ chains. Each chain is calculated by applying the composition
of these functions $M$ times. The beginning and the end of
each chain are stored in the table.

To recover the preimage of a given image $y \in \CalH$,
for each $i$ apply $r_{i}\circ H$ at most $M$ times,
until the end of a chain is found or all the posibilities are exhausted. 
If the end of a chain
is found, the chain is generated from the beginning to retrieve the
preimage of $y$.
All the chains with the same reduction function $r_{i}$ form the
table $T_{i}$.

\subsubsection{Hellman Tables with Distinguished End-Points.}
\label{sub:distinguished end-points}

The idea of \cite{inbook:r-82} was
to use some (distinguished) images as chain ends. For instance all
images $y$ such that $y\leq K$. The parameter $K$ is chosen such that the average
chain length is $M$ (from Equation \ref{eq:m}). 

This improvement reduces the number of table lookups to $M$ in the
worst case (instead of $M^{2}$ in the original Hellman tables) and
makes it easier to detect collisions between chains\footnote{Just compare chain
ends when calculating them, and recalculate on collision. If the
chain is too large, assume a cycle was formed and discard the chain}.
These are the tables and values that we use in our implementation.

\subsubsection{Rainbow Tables.}

Instead of using a single reduction function for each chain, 
Rainbow tables use a different one for each step in the
chain \cite{paper:o-03}. In this paper we do not use this approach 
because each of the PIR queries would be made to a table of size $O(M^2)$,
instead of a table of size $O(M)$ 
(as in Hellman tables with distinguished end-points).
Section~\ref{sec:comparison} completes the comparison between the
two approaches.

\subsection{Password Cracker Routines}

\begin{figure} [ht]
\vspace{-0.5 cm}
\begin{algorithm}[H]
  \caption{Crack password routine}
  \label{algo:password_cracker_routines}%
  \DontPrintSemicolon

  \KwIn{hash}

  \KwOut{password or not found message}

  ends $\gets$ empty\_list() \;
  index $\gets 0$ \;

  \While{\emph{index $< M$}}{
    ends.append(calculate\_end\_point(redfun(index),hash)) \;
	}
  starts $\gets$ empty\_list() \;
  index $\gets 0 $ \;

  \ForEach{\emph{end $\in$ ends}}{

    starts.append( fetch\_start\_PIR(end, index) ) \;
    index $\gets$ index $+ 1$ \;
	}
  index $\gets 0 $ \;

  \ForEach{\emph{start $\in$ starts}}{

    \If{\emph{start != START\_NOT\_FOUND}}{

      solution $\gets$ find\_preimage(start, redfun(index), hash) \;

      \If{\emph{solution}}{
        return solution \;
      } 
		}
    index $\gets$ index $+ 1$ \;
	}
	
  return PREIMAGE\_NOT\_FOUND
\end{algorithm}
\vspace{-0.5 cm}
\end{figure}

Algorithm \ref{algo:password_cracker_routines} shows the routines used by 
the password cracker (on the client's side). It first calculates all the posible
chain buckets for all the $M$ tables on the server for the hash being reversed.
Then, it looks up the buckets on the table provider (lines 7-9), which returns the end of
the chain associated with this bucket and the beginning of the chain. Finally,
using this information, it searches through the matching chains and finds the hash
preimage.

\begin{figure} [ht]
\vspace{-0.5 cm}
\begin{algorithm}[H]
  \caption{fetch\_start\_PIR routine}
  \label{algo:fetch_start}%
  \DontPrintSemicolon

  \KwIn{end, index}

  \KwOut{fetched start}

  pir\_db $\gets$ select\_pir\_db(index) \;

  bucket\_idx $\gets$ bucket\_for(end) \;

  fetched\_start, fetched\_end $\gets$ fetch\_bucket(end, pir\_db) \;

  \If{\emph{fetched\_end != end}}{
    return START\_NOT\_FOUND \;
  }
  return fetched\_start \;
\end{algorithm}

\vspace{-0.5 cm}
\end{figure}

Algorithm \ref{algo:fetch_start} shows how the PIR database is used. Each query
to the table provider is aimed at a different PIR database,
and each database corresponds to a Hellman table. 
After the database is selected, the client calculates the associated
bucket for the distinguished end-point. Finally, the client queries the database, which
returns a pair $(fetched\_start, fetched\_end)$. If the $fetched\_end$ is
the same as the end queried, then the start is returned. If not, then the chain
is not in the table queried.

\section{Complexity Calculation} \label{sec:complexity}

\subsection{PIR Protocols} \label{sec:PIR-details}

In this section, we provide the details of some single-database PIR schemes.
Recall from Section \ref{sec:PIR} that 
$O_{C}$ denotes the client processing complexity,
$O_{S}$ the server processing complexity and 
$O_{T}$ the transfer complexity.

\subsubsection{Naive PIR.}

A very simple scheme for private information retrieval is to simply
send the entire database to the client. The problem with this solution
is that $O_{T}=O(n)$ where $n$ is the size of the database. Also
$O_{S}=O_{C}=O(n)$.

\subsubsection{Classic PIR.}

This is the first single-database PIR
scheme published \cite{paper:pir_classic}. We will analyze the non-recursive
version defined in the first part of the paper. It is based on the
\emph{Quadratic Residuosity Assumption} (QRA) \cite{paper:bbs-86,paper:bc-86,paper:bdmp-91,paper:gm-84}.
The client sends $\sqrt{n}$ numbers to the server where all the
numbers but one are squares. The server, for each row in the database,
multiplies the number if and only if it is ``1'' and
answers with the $\sqrt{n}$ multiplication results. Then the client
chooses the result of interest and calculates whether it is a square.
The requested bit is $0$ if it is a square (and $1$ if not).

Using this scheme the complexities are $O_{S}=n$ and $O_{T}=O_{C}=O(\sqrt{n})$.

\subsubsection{Fast PIR.}

This scheme was published in 2007, and is based
on the hidden-lattice problem \cite{paper:ag-07}. 
The authors claim in \cite{paper:a-08} to have implemented the
first practical computational PIR scheme,
processing in the server 2 Gbits/s. 

In this scheme, $n = O(e \cdot \log(e))$, where $n$ is the number of bits
in the database and $e$ is the number of entries in the database.
Letting $f^{-1}$ be the inverse of $f(x) = x \cdot \log(x)$,
we have that $O_{S}=n$, $O_{T}=O_{C}=O(f^{-1}(n))$.
Even though this scheme has worse client and transmission complexity,
its authors claim that it works in practice better than classic PIR.

\subsection{Password Cracker (the Client)} 
\label{sec:password-cracker}

When a password cracker attempts to break a password, it sends $M$
queries to the table provider and for each response it iterates through
the entire chain (of size $M$) looking for the preimage. Assuming that a step in
the chain can be calculated in $O(1)$, the complexity for each query
is $O_{C}+M$, 
where $O_{C}$ is the client side complexity of a single query in
the PIR scheme used.
The time complexity to find a preimage is thus $M \times (O_{C}+M)$.

We calculate below the complexity $O_{C}$ for the 3 PIR schemes outlined in the
previous section. We denote $S$ the size of the database being
queried (in our solution $S = \beta M$).
\begin{description}
\item [{Naive:}] The complexity $O_{C}$ is $O(S)$.
\item [{Classic:}] The complexity $O_{C}$ is $O(\sqrt{S})$.
\item [{Fast:}] The complexity $O_{C}$ is $O(f^{-1}(S))$, where $f(x) = x \cdot \log(x)$.
\end{description}
As a consequence, in the three cases, the total complexity for the password cracker is 
dominated by $O(M^{2})$.
It is interesting to note that the complexity is not affected by
the fact that we query the database through a PIR protocol.

\subsection{Table Provider (the Server)}

For the server, there are two stages: (i) generating the tables, and
(ii) answering queries.
During the second stage, when the client attempts to crack a password, 
it makes $M$ queries.
Therefore, on the server side, the time complexity of a password crack attempt 
is $O(M \times O_{S})$
where $O_{S}$ is the server processing complexity of a single query in
the PIR scheme used. Recall from the previous sections
that $O_{S} = O(\beta M) = O(M)$, so the total complexity is $O(M^2)$.

Let us analyze now the complexity of generating and storing the tables.
We prove below that the table calculation is bounded in space by $\beta M^{2}$ 
and in time by $\beta'M^{3}$ where $\beta$ and $\beta'$ do not depend on
$M$.

\subsubsection{Expected Number of Chain Calculations.
\label{sub:number-of-chains-calculation}}

When the Hellman tables are calculated and stored in the table provider,
our design decision was to store each Hellman table in a closed hash table,
and to discard the chains when a collision is detected
(see Algorithm \ref{algo:tables_calculation}).
Assuming that the hash function we are trying to invert is cryptographically
strong, it is safe to assume that the all the buckets are equiprobable
and independent for a chain.

Say that we are calculating a table with $M$ entries in a hash
table of size $k=\beta M$ (where $\beta>1$) and that there are $i$
slots occupied with already calculated chains. The probability of
a collision between the next chain to be calculated and the ones already
stored is $\nicefrac{i}{k}$. Therefore, the number of chains to be
calculated follows a geometric probability distribution with expected
value $\nicefrac{k}{k-i}$. So the expected number of chain calculations
for each table ($E_{C}$) is calculated as follows:
\[
E_{C} = \sum_{i=1}^{M-1}\frac{k}{k-i}=k\sum_{i=1}^{M-1}\frac{1}{k-i}
\]
Let $k=\beta M$, we obtain
\[
E_{C} = \beta M\sum_{i=1}^{M-1}\frac{1}{\beta M-i}
\]
Let $j=\beta M-i$, we get
\[
E_{C} = \beta M\sum_{j=\beta M-1}^{\beta M-M+1}\frac{1}{j}=\beta M\left[\sum_{j=1}^{\beta M-1}\frac{1}{j}-\sum_{j=1}^{\beta M-M+1}\frac{1}{j}\right]
\]
By approximating the harmonic series partial sums as $\log(n)$:
$$
E_{C} \simeq \beta M\left(\log(\beta M-1) - \log(\beta M-M+1)\right) 
$$
Then $\forall \, M > C, \; \exists \, c>1$ such that:
\begin{align*}
E_{C} \; & \le \; \beta M\left(\log(\beta M) - \log(c\,(\beta-1)M)\right)  \\
 &= \; \beta M\left(\log(\beta)+\log(M)-\log(\beta-1)-\log(M)-\log(c\right)) \\
 &= \; \beta M\left(\log(\beta)-\log(\beta-1)-\log(c\right))
\end{align*}
Proving that $\forall \, \beta > 1, \; \exists \, \beta'$ 
such that $E_{C} \leq \beta'M$. 
This means that when $M$ increases, if the ratio of unused space
is kept constant in the closed hash, the time used to calculate the
table increases linearly or sublinearly.

\subsection{Transfer Complexity}

The transfer complexity of a password crack attempt is $O(M\times O_{T})$
where $O_{T}$ is the transfer complexity of a single query in the
PIR scheme used. We calculate this complexity for the 3 PIR schemes
outlined in Section~\ref{sec:PIR-details}.
Again we denote $S$ the size of the database being queried (here $S = O(M)$).
\begin{description}
\item [{Naive:}] The time complexity for a single query is $O(S)$. 
So the total complexity for the transfer is $O(M^{2})$.
\item [{Classic:}] The time complexity for a single query is $O(\sqrt{S})$.
So the total complexity for the transfer is $O(M^{3/2})$.
\item [{Fast:}] The time complexity for a single query is $O(f^{-1}(S))$,
where $f(x) = x \cdot \log(x)$. So the total complexity for the
transfer is $O(M \cdot f^{-1}(M))$.
\end{description}

\subsection{Comparison with Other Approaches} \label{sec:comparison}

\subsubsection{Brute-force.}

If the password cracker brute forces the password she will not have
transfer costs but she will have to iterate through the entire preimage
space. So the processing costs for the password cracker will be in
the order of $M^{3}$.

\subsubsection{Rainbow Tables.}

This approach is much slower because in order to use a Rainbow table
each query must be made against $M^{2}$ entry points, instead of
the $M$ entry points for each table when using Hellman tables with
distinguished end-points, and each PIR query resource usage depends
on the number of entries in the table (see Section~\ref{sec:PIR-details}).

\section{Implementation and Experimental Results}
\label{sec:experimental-results}

We have made a prototype that implements the solution described in this paper with
the naive and classic PIR schemes. 
Namely we implemented hash reversing for the hash function MD5. 
We considered passwords using an alphabet $\CalA$ of 6 letters,
and passwords lengths of 4, 5 and 6 characters. We also choose $\beta = 4$.

We focused our performance evaluation on the following parameters:
(i) the length $\ell$ of the passwords ($ 4 \leq \ell \leq 6$), 
and  (ii) the desired probability $\alpha$ of reversing the 
hashes ($0.4 \leq \alpha \leq 0.9$).
By varying $\ell$ and $\alpha$, we obtain different values for $M$ 
(the size of the database). Recall from Section~\ref{sec:hash-reversing}
that $M$ is computed as
$$
M = -\sqrt[3]{ \ln(1-\alpha) \cdot N}
$$
where $N$ is the total number of passwords considered.
In this case $N = | \CalA | ^{\ell} $.
More concretely, $N = 1296$ for $\ell=4$, $N = 7776$ for $\ell = 5$,
and $N = 46656$ for $\ell = 6$.

\begin{figure}[ht]
\centering
\includegraphics[width=0.9 \linewidth]{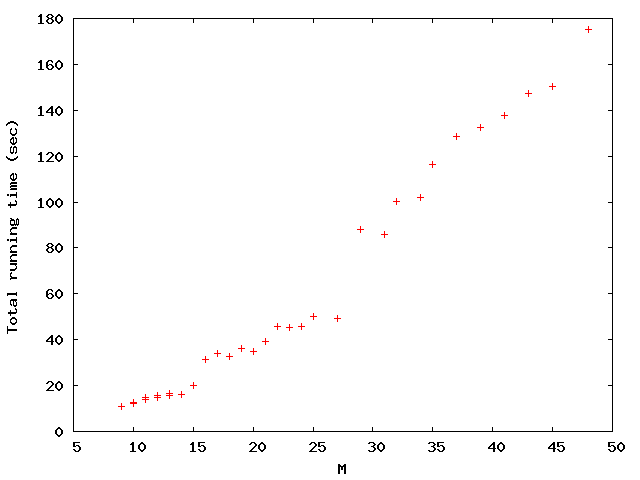}
\caption{Total running time as a function of $M$ (the size of the database).}
\label{fig:results}
\end{figure}

Figure~\ref{fig:results} shows the total runtime of the solution 
as a function of $M$ (each point represents the runtime 
to crack 100 passwords).
The tests were performed on a Linux virtual machine. The guest
has one CPU and 1.5 Gb of RAM. The host is also running Linux, it has 8
Gb of RAM and an Intel Xeon CPU @ 3.30GHz.
The figure shows that the running time grows with $M$, but
also depends on $\ell$ (the points are grouped in three ``clusters''
corresponding to $\ell = 4,5,6$).
The results obtained are coherent with the estimated complexities.
However we think that more extensive testing is needed to validate
this approach.

We believe that our implementation still has lots of room for improvement.
To begin with, the implementation is in pure Python,
and thus much slower than its equivalent in C++ for example.
Because of that, we are using small primes (of 20 bits) 
to implement classic PIR.
We also think that the reduction function could be optimized. 
We discuss further ideas for improvement in the next section.

\section{Conclusion and Future Steps} \label{sec:conclusion}

In this paper we tackle a problem that, as far as we know,
has not been studied before. We propose a first solution
for building an ``oblivious password cracking server'', 
that preserves the privacy of the queries made to the server.
Even though the expected performance of this password cracker
is still not sufficient to use it in real life scenarios, 
this scheme provides perfect privacy,
and is better (in terms of complexity) than other potential approaches
that we considered.

A natural future step for this project is 
to benefit from the recent advances in PIR research.
For instance, Ian Goldberg and his group
at the University of Waterloo have developed a PIR library in C++
(called Percy++), which implements protocols based on \cite{Goldberg07}.
Recently, they have used it to develop
PIR protocols for electronic commerce that they claim to 
be ``practical'' (presented in 2011 at the ACM CCS conference
\cite{HenOluGol11}).
We believe that using Percy++ the results obtained here could
be greatly improved.

Another promising direction is to exploit the fact that
the solution proposed is inherently parallelizable. 
Each PIR database could run in a different host, 
and the client work could also be trivially parallelized.
An additional improvement would be to run the password cracking server ``in the cloud'',
with the possibility of adding processors on demand.
To conclude, we hope that the first step presented here will inspire other researchers as well.

\bibliography{../paper}{}
\bibliographystyle{plain}

\end{document}